\documentclass[
showkeys,	%	Control display of Keywords: line.
reprint,	%	Two-column formatting.
amsmath,	%	Load amsmath package.- AMS-LATEX features.
amssymb,	%	Load amssymb package.-  AMS symbols.
amsfonts,	%	Load amsfonts package.-  AMS font support.
aps,		%	American Physical Society styling.
pre,%
groupedaddress,% Group authors with same affiliations.
superscriptaddress,% Associate authors with 			affiliations via superscript numbers. Appropriate for collaborations or if several authors share some, but not all, affiliations.
a4paper,%
nofootinbib,%	Control location of footnotes.
eprint,		%	arXiv e-print identifiers in bibliography.
final,		%	Don’t mark overfull lines.
preprintnumbers, % Control display of preprint numbers given by \preprint command.
]{revtex4-2}

\usepackage[tbtags]{mathtools}
\usepackage{amsmath,amsfonts,mathdots,amssymb,yfonts,calc}
\usepackage{graphicx,graphics,xcolor}
\usepackage{subfigure}

\usepackage{natbib}
\usepackage{booktabs}
\usepackage{siunitx}
\usepackage{float}
\usepackage{tabularray}
\usepackage{multirow}
\usepackage{physics}
\usepackage{dcolumn}% Align table columns on decimal point

\usepackage{hyperref}%reference back to page
\usepackage{natbib}
\usepackage{booktabs}
\hypersetup{
	colorlinks=true,
	linkcolor=blue,
	filecolor=magenta,      
	urlcolor=blue!80,
	citecolor=cyan,
	linktocpage=true,
	breaklinks=false,
}

\begin{document}
\title{Noise Reinstates Collapsed Populations: Stochastic Reversal of Deterministic Extinction}
%*******************************************************
\author{Vinesh Vijayan}
\email[]{vinesh.physics@rathinam.in}
\affiliation{Department of Science and Humanities, Rathinam Technical Campus, Coimbatore, India -641021}
%&&&&&&&&&&&&&&&&&&&&&&&&&&&&&&&&&&&&&777
\author{B Priyadharshini}
%\surname{Pranaya Pratik},\surname{Das}
%\email{priyadharshini.sh@rathinam.in}
\affiliation{Department of Science and Humanities, Rathinam Technical Campus, Coimbatore, India -641021}
%$$$$$$$$$$$$$$$$$$$$$$$$$$$$$$$$$$$$$$
\author{R Sathish Kumar}
%\email[]{sathishkumar.maths@rathinam.in}
\affiliation{Department of Science and Humanities, Rathinam Technical Campus, Coimbatore, India -641021}
%**************************************************
\author{G Janaki}
%\email[]{janaki.sh@rathinam.in}
\affiliation{Department of Science and Humanities, Rathinam Technical Campus, Coimbatore, India -641021}
%**********************************************************

\date{\today}

\begin{abstract}
Conventional wisdom suggests that environmental noise drives populations toward extinction. In contrast, we report a paradoxical phenomenon in which stochasticity reverses a deterministic tipping point, thereby preventing collapse. Using a hybrid model that integrates logistic growth with a density-triggered sigmoidal collapse, we uncover a striking reversal: deterministic fragility on one side, and stochastic rescue under weak noise on the other. Our analysis demonstrates that noise disrupts the convergence of deterministic trajectories toward extinction by altering the phase space topology, enabling back-transitions to viable states. This mechanism gives rise to noise-induced metastability and reveals a form of stochastic robustness not captured by deterministic models. These findings suggest that natural fluctuations can serve as a stabilizing force in complex systems, offering a compelling counter-narrative to classical models in ecology, epidemiology, and beyond. We advocate for a re-evaluation of stabilization strategies, emphasizing the constructive role of stochasticity in averting population collapse.
\end{abstract}

%\maketitle must follow title, authors, abstract, and keyword
\maketitle
%\tableofcontents
\pagestyle{plain}

\section{\label{sc1}Introduction}
Discrete-time dynamical systems are powerful tools for modelling because they exhibit bifurcations as system parameters vary, leading to qualitative changes in dynamics \cite{may1976simple, hilborn2000chaos}. The logistic and hybrid maps long served as paradigms for studying period doubling, bifurcation, chaos, and extinction phenomena in complex systems\cite{may1976simple, ForgostonMoore2017, MatsumotoTsuda1983}.  A key advantage of using these discrete-time models are their low data requirement; they are especially suitable when system states—such as population counts or gene expressions—are sampled at regular intervals \cite{kuehn2011mathematical}.
Discrete-time systems are also computationally tractable, enabling high-resolution simulations and extensive parameter sweeps. In scenarios where continuous-time modelling is either computationally infeasible or empirically unwarranted, discrete maps bridge the gap between theoretical understanding and empirical observation \cite{guttal2008changing, wissel1984universal}. But real world systems are seldom deterministic, intrinsic or external perturbations significantly alter stability, persistence and transition between states dramatically \cite{BiYang2015, Majda2021, Spagnolo2003}. An exceptional tool for understanding noise-driven transitions is the quasi-potential landscape \cite{Boettiger2012, Zhang2014}. 

Early Warning Signals (EWS) are statistical indicators that can predict these transitions before they occur, providing crucial insights into approaching tipping points \cite{scheffer2009early, dakos2012methods, boettiger2013early}. EWS metrics such as rising autocorrelation, increased variance, and critical slowing down can be computed from time series data to anticipate upcoming transitions \cite{scheffer2009early, dakos2008slowing}. In many real-world systems, critical transitions are not only caused by deterministic bifurcations but also by stochastic perturbations. EWS have shown promise in predicting such noise-induced transitions in hybrid stochastic maps \cite{ashwin2012tipping, ghil2008climate}. Monitoring EWS thus offers a window into the resilience of systems, which is vital in contexts where pre-emptive interventions could prevent collapse or unwanted shifts \cite{carpenter2011early}. Therefore, incorporating EWS into discrete systems not only deepens our theoretical insights but also enhances their practical utility in biology, ecology, and physics.

In the literature, threshold-driven population collapses are typically modelled using either piecewise functions~\cite{carpenter2011} or ordinary differential equation (ODE) bifurcations~\cite{scheffer2001}. It has been observed that the introduction of noise amplifies extinction risk~\cite{lande1993}. Sigmoidal terms~\cite{holling1959} are commonly used to represent saturating predation, but they are not traditionally employed to model intrinsic collapse dynamics in populations. Furthermore, it is often found that population recovery after collapse requires external immigration~\cite{gonzalez2002} In this work, we demonstrate that these distinct paradigms can be unified within a single framework: a discrete-time map. By incorporating a sigmoidal collapse term and stochasticity, our model exhibits population recovery and stabilization without the need for external input, in contrast to previous findings~\cite{lande1993}. This approach provides a novel bridge between nonlinear dynamics and stochastic resilience in ecological systems.

May’s seminal work demonstrated that even simple recurrence relations, such as the
logistic map, can exhibit a period-doubling route to chaos. This foundational insight has
since catalysed extensive research into discrete-time dynamical systems. The logistic map,
in particular, has proven to be a powerful framework for understanding the emergence of
complexity and chaotic behaviour in a wide range of systems \cite{may1976simple, hilborn2000chaos}. In a number of influential studies, rational activation terms have been shown to effectively model threshold-like behaviour, with direct implications for gene regulatory mechanisms\cite{holling1959}. Such formulations
capture essential features of genetic feedback, including stochastic switching in
self-regulatory genes—a hallmark of gene expression dynamics. Furthermore, several investigations have demonstrated that bistability arising from rational activation can serve
as a fundamental mechanism for modelling gene regulatory networks, thereby reinforcing
the practical significance and applicability of this approach in the study of systems
biology\cite{Alon2006, Zhang2014}. Real-world systems often exhibit growth dynamics coupled with nonlinear inhibitory feedback, with stochastic influences acting as external perturbations. 

In the light of above observations, we propose and analyse a hybrid nonlinear map that integrates logistic growth with a sigmoidal saturation term, representing an internal inhibitory mechanism.
This framework enables a focused investigation into the system’s dynamics under stochastic forcing, highlighting how the interplay between growth and inhibition shapes the underlying stability landscape. EWS measures are used to identify the tipping points and potential energy landscape will provide a clear picture of transitions under noise. Our numerical analyses reveal rich bifurcation behaviour, including noise-induced transitions, positioning the system
as a valuable testbed for evaluating the performance and robustness of early warning
signals (EWS) beyond conventional logistic or bistable paradigms. Key EWS indicators—
such as increasing variance, autocorrelation, and skewness—are assessed in the vicinity of critical transitions. The findings provide new insights into the resilience and recovery characteristics of complex adaptive systems, with significant
implications for understanding ecological and climate-related tipping points.

The paper is organized as follows. In Section\ref{sec2}, we present the mathematical formulation of the model. Section\ref{sec3} discusses the deterministic dynamics of the system, while Section\ref{sec4} explores its stochastic behaviour. In Section\ref{sec5}, we present the key findings and discuss their significance in relation to existing literature. Finally, we conclude the paper in Section\ref{sec6}.

\section{\label{sec2}Mathematical Modelling}
\subsection{Model Description}
In general, we consider a system governed by a discrete-time map of the form

\begin{equation}
x_{n+1} = f(x_n; \alpha, \beta)
\label{eq1}
\end{equation}

where  \( x_n \in \mathbb{R} \) represents the state variable (e.g., population density in the logistic map), and \( \alpha \) and \( \beta \) are system parameters that control the dynamics. 
The \textit{stochastic version} of the model includes additive noise and is given by:

\begin{equation}
x_{n+1} = f(x_n; \alpha, \beta) + \sigma \eta_n
\label{eq2}
\end{equation}

here \( \sigma \) denotes the noise strength, and \( \eta_n \sim \mathcal{N}(0,1) \) is a standard Gaussian white noise term. This formulation captures the influence of random fluctuations on the system’s evolution over time. We choose $f(x_n; \alpha, \beta)$ with a hybrid nature-a discrete time map of the following form:

\begin{equation}
x_{n+1} = r x_n (1 - x_n) - \frac{x_n^2}{x_n^2 + a^2} + \sigma \eta_n
\label{eq3}
\end{equation}

The parameters \( \alpha = r \) and \( \beta = a \) control the strength of the logistic growth and the sigmoidal inhibition term, respectively. \( r > 0 \) is the logistic growth rate, and \( a > 0 \) determines the steepness and threshold of the inhibitory term.

The logistic growth term with carrying capacity forms the core of the model. An additional inhibition term, functioning as a smoothed step function, introduces a critical layer of nonlinearity. When the population density \( x_n \) is small relative to the threshold parameter \( a \), the inhibitory effect is minimal. However, as \( x_n \) increases and exceeds \( a \), the inhibition term asymptotically approaches \(-1\), generating strong negative feedback. At \( x_n = a \), the function exhibits a smooth threshold transition—a key mechanism underlying collapse dynamics. This threshold-driven behavior can lead to abrupt population declines once the critical density \( a \) is surpassed, effectively capturing tipping-point phenomena observed in real-world systems such as algal blooms and tumor growth. The model thus offers a powerful framework for understanding sudden population collapses and state transitions triggered by high-density thresholds. Furthermore, the incorporation of stochastic noise makes the model more realistic for ecological applications, enabling the study of fluctuations and resilience near critical transition points.

\subsection{Fixed Points and Stability}
To find out the fixed poits and stability consider the discrete-time map defined by:

\begin{equation}
x_{n+1} = r x_n (1 - x_n) - \frac{x_n^2}{x_n^2 + a^2}
\label{eqn4}
\end{equation}

Fixed points \( x^* \) of the system satisfy:

\begin{equation}
x^* = r x^* (1 - x^*) - \frac{(x^*)^2}{(x^*)^2 + a^2}
\label{eqn5}
\end{equation}

Bringing all terms to one side yields:

\begin{equation}
0 = r x^* (1 - x^*) - x^* - \frac{(x^*)^2}{(x^*)^2 + a^2}
\label{eqn6}
\end{equation}

Equation~\eqref{eqn6} is a transcendental equation and generally cannot be solved analytically in closed form. However, the fixed points \( x^* \) can be located numerically or graphically for given values of the parameters \( r \) and \( a \).
To analyse the stability of the fixed points, we define the map function:

\begin{equation}
f(x) = r x (1 - x) - \frac{x^2}{x^2 + a^2}
\label{eqn7}
\end{equation}

The stability of a fixed point \( x^* \) depends on the derivative \( f'(x) \), given by:

\begin{equation}
f'(x) = r(1 - 2x) - \frac{d}{dx} \left( \frac{x^2}{x^2 + a^2} \right)
\label{eqn8}
\end{equation}

We compute the derivative of the inhibitory term:

\begin{align}
\frac{d}{dx} \left( \frac{x^2}{x^2 + a^2} \right) 
&= \frac{2x(x^2 + a^2) - x^2(2x)}{(x^2 + a^2)^2} \label{eqn9} \\
&= \frac{2x a^2}{(x^2 + a^2)^2}
\label{eqn10}
\end{align}

Substituting back into Equation~\eqref{eqn8}, the full derivative becomes:

\begin{equation}
f'(x) = r(1 - 2x) - \frac{2x a^2}{(x^2 + a^2)^2}
\label{eqn11}
\end{equation}

The stability condition for a fixed point \( x^* \) is:

\begin{itemize}
  \item \( |f'(x^*)| < 1 \): \quad Stable fixed point
  \item \( |f'(x^*)| > 1 \): \quad Unstable fixed point
  \item \( f'(x^*) = -1 \): \quad Candidate for period-doubling bifurcation
\end{itemize}
\section{\label{sec3}Deterministic Dynamics}
\subsubsection{Lyapunov Exponent Landscape}

We numerically evaluate the largest Lyapunov exponent \( \lambda \) over the \((r, a)\) parameter space to classify the system's dynamics into fixed points, limit cycles, and chaotic regimes. The Lyapunov exponent is computed as:

\begin{align}
\lambda &= \lim_{N \to \infty} \frac{1}{N} \sum_{n=1}^{N} \ln \left| f'(x_n) \right| \nonumber \\
        &= \lim_{N \to \infty} \frac{1}{N} \sum_{n=1}^{N} \ln \left| r(1 - 2x_n) - \frac{2x_n a^2}{(x_n^2 + a^2)^2} \right|
\label{eqn12}
\end{align}

\begin{figure}[htbp]
    \centering
    \includegraphics[width=0.5\textwidth]{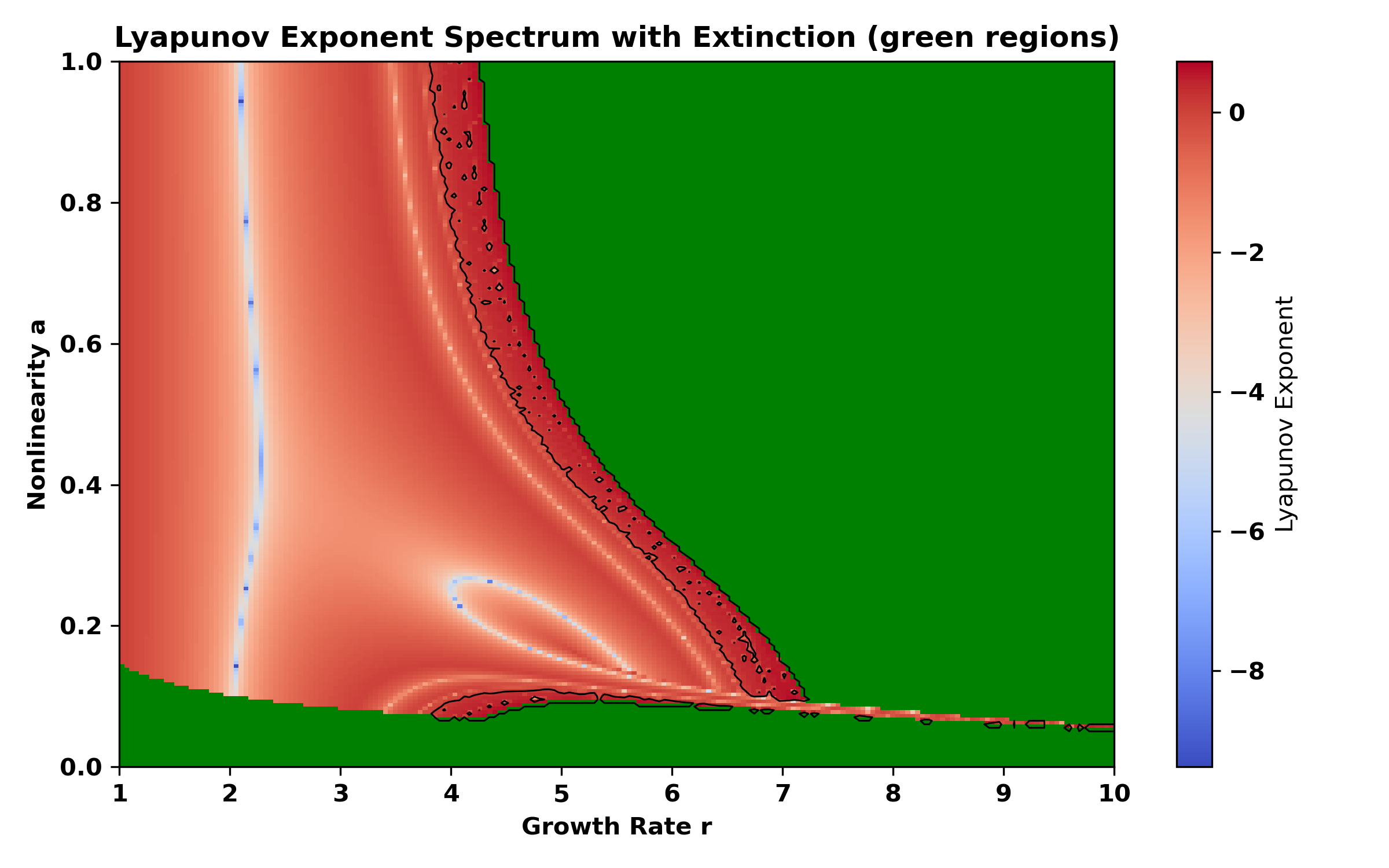}
    \caption{Lyapunov exponent spectrum over the $(r, a)$ parameter space showing regions of stability and chaos. The color encodes the magnitude of the largest Lyapunov exponent
The black contour line demarcates the transition between regular and chaotic regimes. Green color highlights region of extinctions}
    \label{fig1}
\end{figure}

\begin{figure}[htbp]
    \centering
    \includegraphics[width=0.5\textwidth]{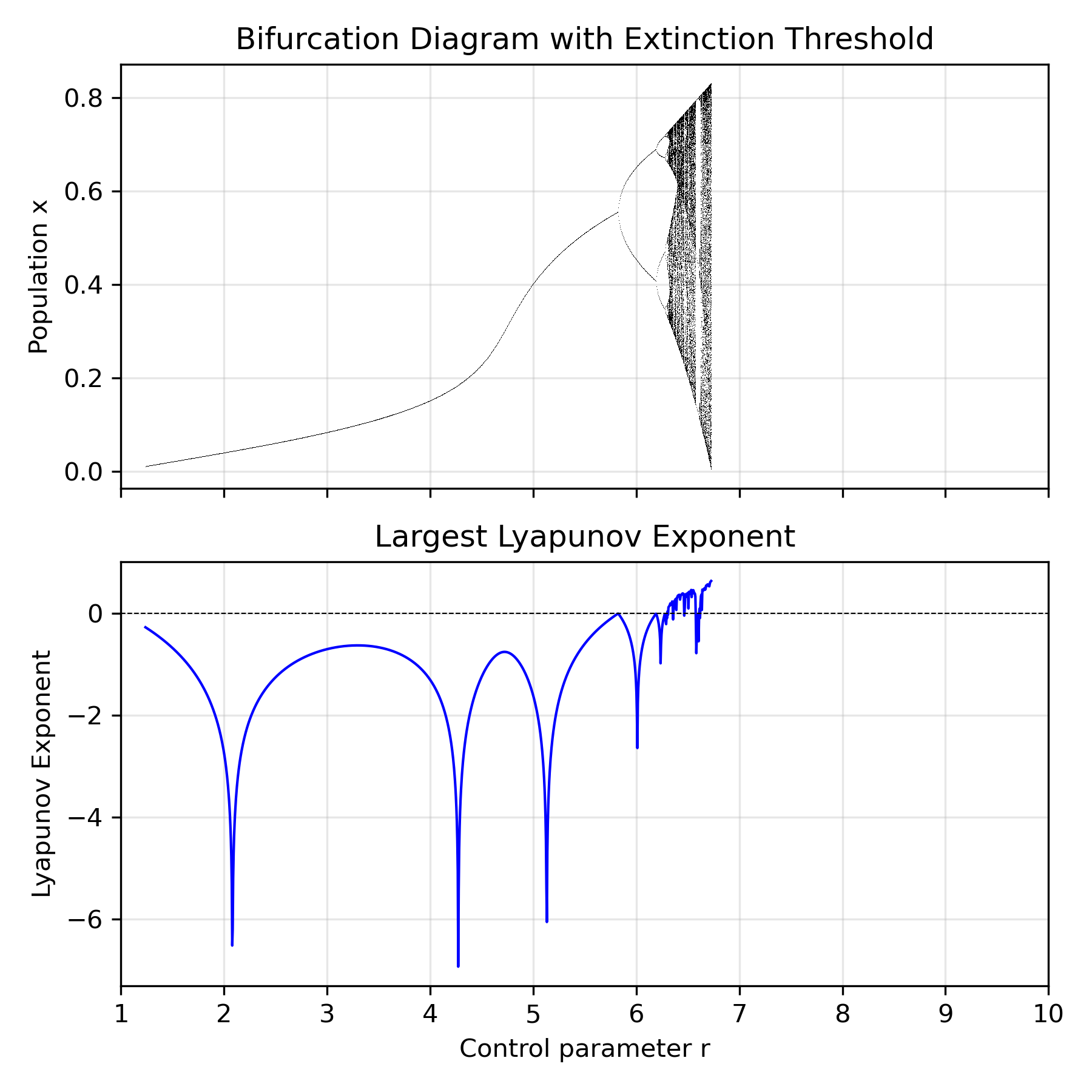}
    \caption{Bifurcation diagram (top) and corresponding largest Lyapunov exponent
(bottom) for a population model with an extinction threshold, plotted as a function of
the control parameter r. The bifurcation diagram illustrates how long-term population
values change with r, revealing fixed points, periodic orbits, and chaotic behavior, while
accounting for extinction when the population drops below a critical threshold. The
Lyapunov exponent plot quantifies the stability of these dynamics, with negative values
indicating stable behavior and positive values indicating chaos.}
    \label{fig2}
\end{figure}

When the parameter value \( a \) is small (see FIG~\ref{fig1}), the system exhibits classical routes to chaos as the control parameter \( r \) is varied. As \( r \) increases from a low value, the dynamics of the hybrid map transition from a stable fixed point to periodic oscillations, and eventually to chaos—consistent with the traditional framework of chaos theory. However, as \( a \) increases, there are significant shifts in both the location and extent of chaotic regions within the parameter space. This highlights the substantial influence of the nonlinear inhibitory feedback introduced by the sigmoidal term governed by \( a \). The Lyapunov spectrum, a powerful analytical tool, provides a clear and comprehensive depiction of stability, periodicity, and chaotic behaviour across the parameter space, thereby elucidating the system’s global dynamical structure. Regions in the parameter space shaded in green indicate zones of extinction or collapse($x_n=a$). To explore this phenomenon in greater detail, we fix the value of \( a = 0.2 \) and vary the control parameter \( r \), observing the resulting bifurcation structure (see FIG~\ref{fig2}).

An intriguing behaviour is observed beyond a critical value of \( r \approx 6.7 \), for a fixed value of \( a \), where no sustained attractors appear in the bifurcation diagram. This phenomenon can be attributed to \textit{transient chaos}, which eventually decays into extinction. Mathematically, the system is capable of supporting chaotic dynamics; however, the presence of the second term in the dynamical equation—a component that becomes dominant at low population levels—plays a pivotal role in driving the system toward population collapse. The system’s Lyapunov exponent initially indicates unpredictability and strong sensitivity to initial conditions. Nevertheless, the absence of persistent points in the bifurcation diagram suggests that the population either diverges numerically before any long-term behaviour can emerge, or drops below a critical extinction threshold. This result highlights the paradoxical role of chaos in biological systems: while it may sustain complex dynamics internally, it can also render populations more fragile and prone to extinction. 

\begin{figure}[htbp]
    \centering
    \includegraphics[width=0.45\textwidth]{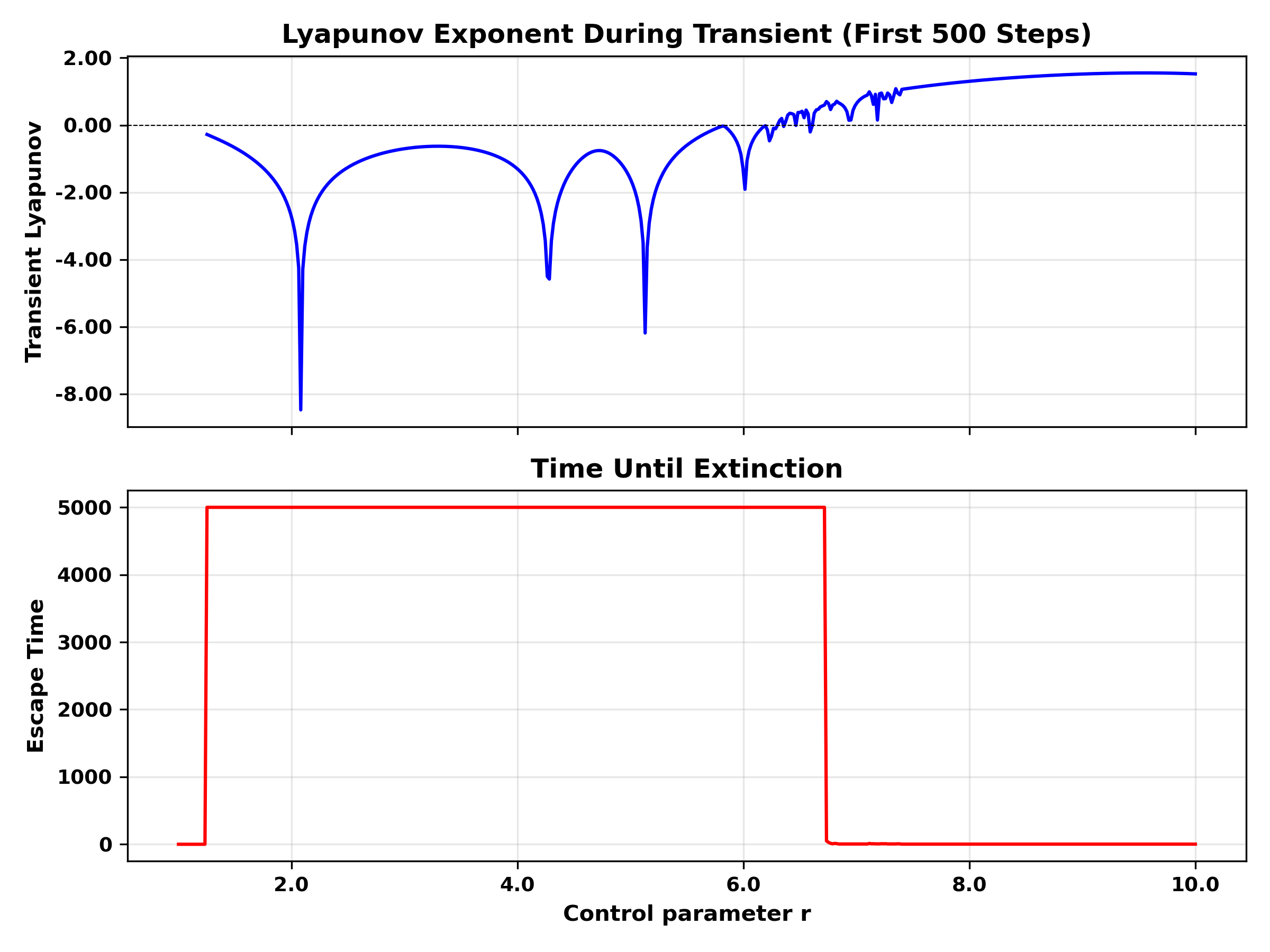}
    \caption{Transient analysis of a dynamical system as a function of the control parameter
r. The top panel shows the transient Lyapunov exponent over the first 500 iterations,
indicating sensitivity to initial conditions and short-term stability. The bottom panel
displays the time until extinction (escape time), revealing ranges of r where the system
persists or collapses quickly.}
    \label{fig3}
\end{figure}

FIG~\ref{fig3} illustrates that at low values of \( r \), the system quickly goes extinct, indicating insufficient growth or inherent instability. This is further supported by negative Lyapunov exponents, which reflect the immediate consequences of such conditions. For intermediate values of \( r \approx 1.5 \) to \( 6.8 \), the system exhibits longer transients and survives for the entire simulation duration, often displaying complex or near-stable dynamics. This behaviour is indicated by varying Lyapunov exponents that approach or cross zero, suggesting that the system is sufficiently stable within this parameter range. 
However, for larger values of \( r \), even though the Lyapunov exponents become positive—signifying chaotic dynamics—the system tends to go extinct rapidly. In these cases, the population crashes before it can settle into any long-term attractor. This underscores the crucial role of transient dynamics in determining the overall behaviour and fate of the system.

\begin{figure*}[htbp]
    \centering
     \includegraphics[width=0.9\textwidth, height=0.9\textheight]{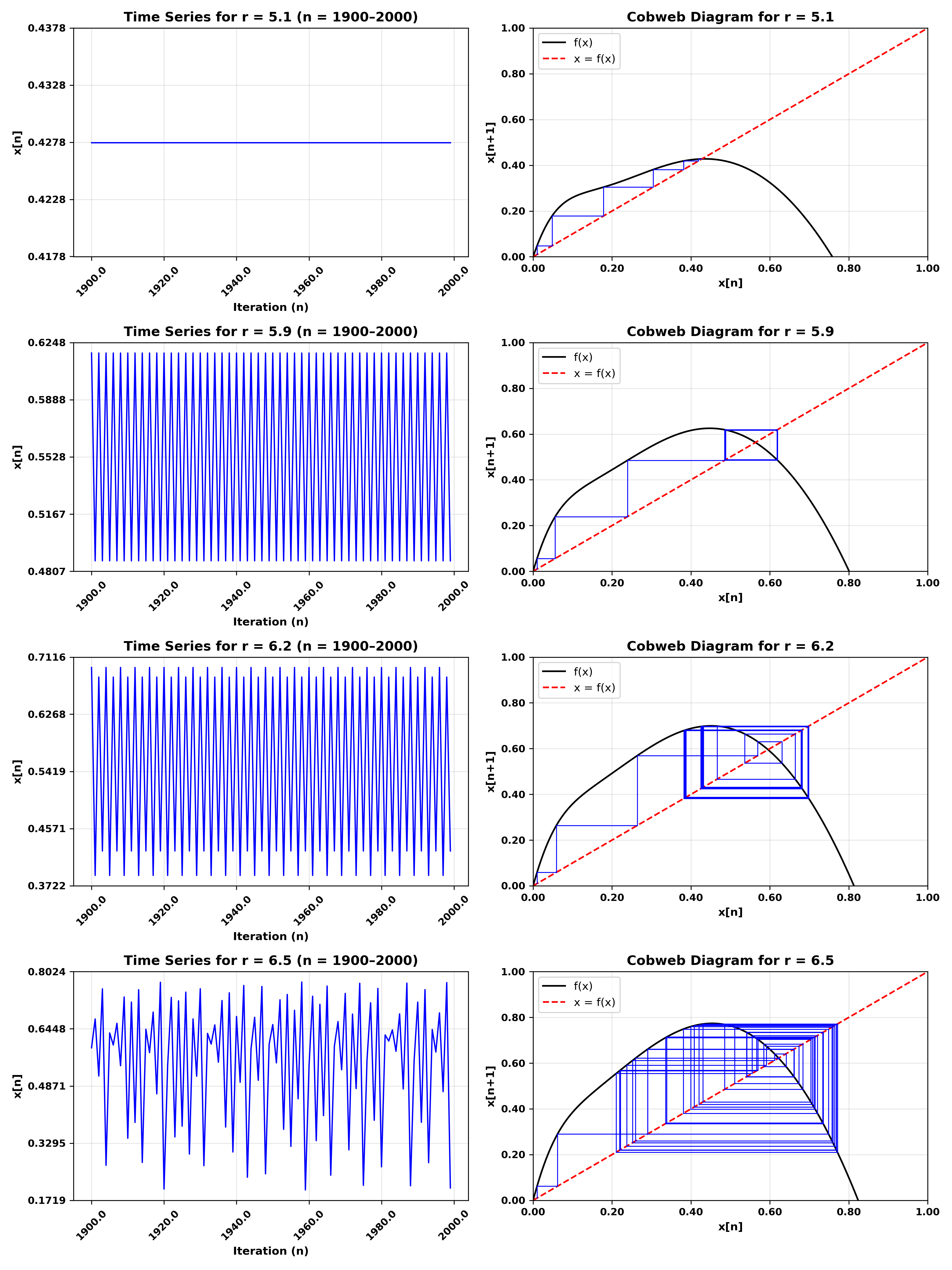}
    \caption{Time series (left) and cobweb diagrams (right) for the hybrid map with different
values of the parameter r: 5.1, 5.9, 6.2, and 6.5. Each row shows how the system
evolves over time (left) and the corresponding cobweb diagram (right) for a specific r.
As r increases, the behavior transitions from fixed points to periodic oscillations, and
eventually to chaotic dynamics.}
    \label{fig4}
\end{figure*}

FIG~\ref{fig4} illustrates the behaviour of the unperturbed hybrid model for various values of the control parameter \( r \), with the inhibition parameter \( a \) fixed at 0.2. The left column presents time series corresponding to the system’s asymptotic states under the specified parameter values, while the right column shows Cobweb plots that graphically represent the function \( f(x) \) and its iterated dynamics. 
These plots provide a clear visualisation of the system’s evolution and offer valuable insight into its underlying dynamical structure. For \( r = 5.1 \), the system rapidly converges to a stable fixed point, as evidenced by the spiral convergence pattern in the Cobweb plot. At \( r = 5.9 \), the Cobweb plot reveals regular, repeating oscillations, indicative of a period-one stable limit cycle. This structured behaviour suggests a predictable and robust system dynamic. As \( r \) increases to 6.2, the Cobweb plot exhibits alternating values, signifying the onset of a period-two limit cycle. When \( r = 6.5 \), the system transitions into a chaotic regime, which is characterised by irregular, dense, and non-repeating trajectories in the Cobweb plot—hallmarks of deterministic chaos.

\subsection{Early Warning Signal Indicators}

From the time series of the dynamical system \( \{x_n\}_{n=1}^{N} \) for a fixed parameter \(r\), the EWS can be calculated using the following statistical measures.

\begin{itemize}
    \item \textbf{Variance}:
    \begin{equation}
    \text{Var}(x) = \frac{1}{N} \sum_{n=1}^{N} (x_n - \bar{x})^2
    \label{eqn13}
    \end{equation}
    where \( \bar{x} = \frac{1}{N} \sum_{n=1}^{N} x_n \) is the sample mean. A sharp increase in variance---a key indicator of critical slowing down---suggests that the system is approaching a tipping point.

    \item \textbf{Lag-1 Autocorrelation (AC)}:
    \begin{equation}
    \text{AC}_1 = \frac{\sum_{n=1}^{N-1} (x_n - \bar{x})(x_{n+1} - \bar{x})}{\sum_{n=1}^{N} (x_n - \bar{x})^2}
    \label{eqn14}
    \end{equation}
The rise in lag-1 autocorrelation indicates that the current state is increasingly influenced by its past, a clear sign of reduced resilience.

    \item \textbf{Skewness}:
    \begin{equation}
    \text{Skew}(x) = \frac{\frac{1}{N} \sum_{n=1}^{N} (x_n - \bar{x})^3}{\left( \frac{1}{N} \sum_{n=1}^{N} (x_n - \bar{x})^2 \right)^{3/2}}
    \label{eqn15}
    \end{equation}
Changes in skewness, which reflect asymmetry in fluctuations, often point to emerging bistability or directional tipping.

    \item \textbf{Detrended Fluctuation Analysis (DFA)}:
    \begin{equation}
    F(s) = \sqrt{ \frac{1}{N} \sum_{k=1}^{N} \left[ Y(k) - y_s(k) \right]^2 }
    \label{eqn16}
    \end{equation}
    where \( Y(k) = \sum_{i=1}^{k} (x_i - \bar{x}) \) is the cumulative sum ("profile") of the time series, and \( y_s(k) \) is the local trend (typically fitted via least squares) in a window of size \( s \). The DFA exponent \( \alpha \) is estimated from the scaling relation:
    \[
    F(s) \sim s^{\alpha}
    \]
    where \( \alpha \) characterises the long-range correlation structure.  A high DFA exponent, indicating long-range memory effects, serves as a strong predictor of impending regime shifts and underscores the urgency of early detection.
\item \textbf{Composite early warning signal}:    
   The product \[ \lambda (1 - \text{AC}_1)\] is an EWS measure that combines Lyapunov exponent and lag-1 autocorrelation functions and is known as composite early warning signal.   A decline in the composite early warning signal \( \lambda (1 - \text{AC}_1) \) suggests the onset of transient chaos and reduced system stability, further emphasizing the importance of this analysis.
\end{itemize}
 
\begin{figure}[htbp]
    \centering
    \includegraphics[width=0.5\textwidth]{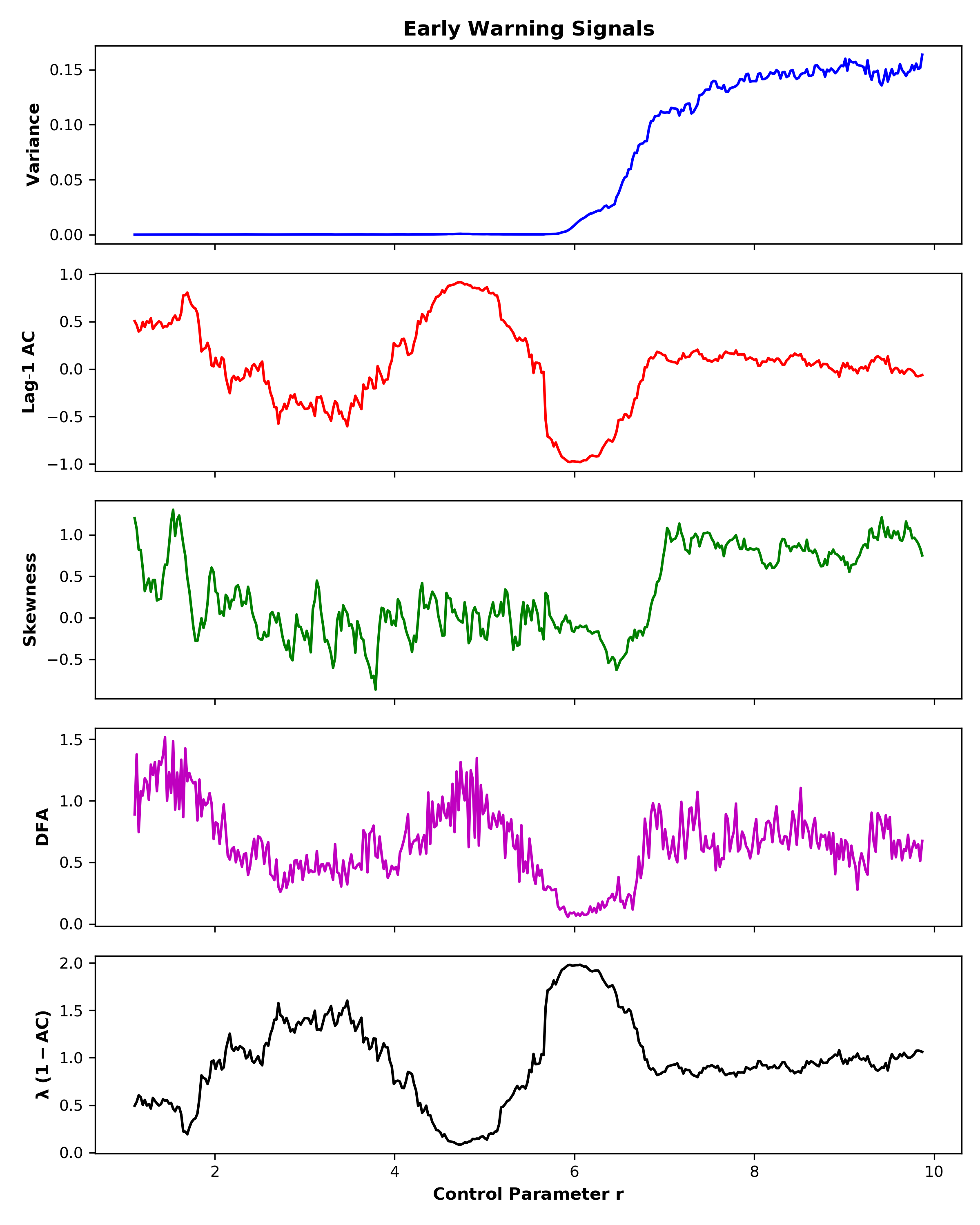}
   \caption{Early warning signals (EWS) computed from the time series of the hybrid dynamical system as the control parameter \( r \) is varied.  These indicators together highlight the system's approach toward a critical transition.}
    \label{fig5}
\end{figure}

From FIG~\ref{fig5}, the variance (\textcolor{blue}{blue}) remains close to zero as the control parameter \( r \) varies, but rises sharply around \( r \approx 5.5 \) and continues to increase, indicating an approaching tipping point. The lag-1 autocorrelation (\textcolor{red}{red}) also peaks in the range \( r \approx 5.5\text{--}6.5 \), highlighting a reduction in system resilience. The skewness (\textcolor{green}{green}) exhibits fluctuations with notable peaks near \( r \approx 5.5 \) and again around \( r \approx 7 \), reflecting the emergence of asymmetry and potential bistability. The DFA exponent (\textcolor{purple}{purple}) shows bursts in the same region, suggesting the presence of regime shifts and long-range correlations. Finally, a dip in the composite early warning signal \( \lambda(1 - \text{AC}_1) \) (\textcolor{black}{black}) coincides with these values, indicating the onset of transient chaos and reduced stability.
\section{\label{sec4}Stochastic Dynamics}
\begin{figure}[htbp]
    \centering
    \includegraphics[width=0.5\textwidth]{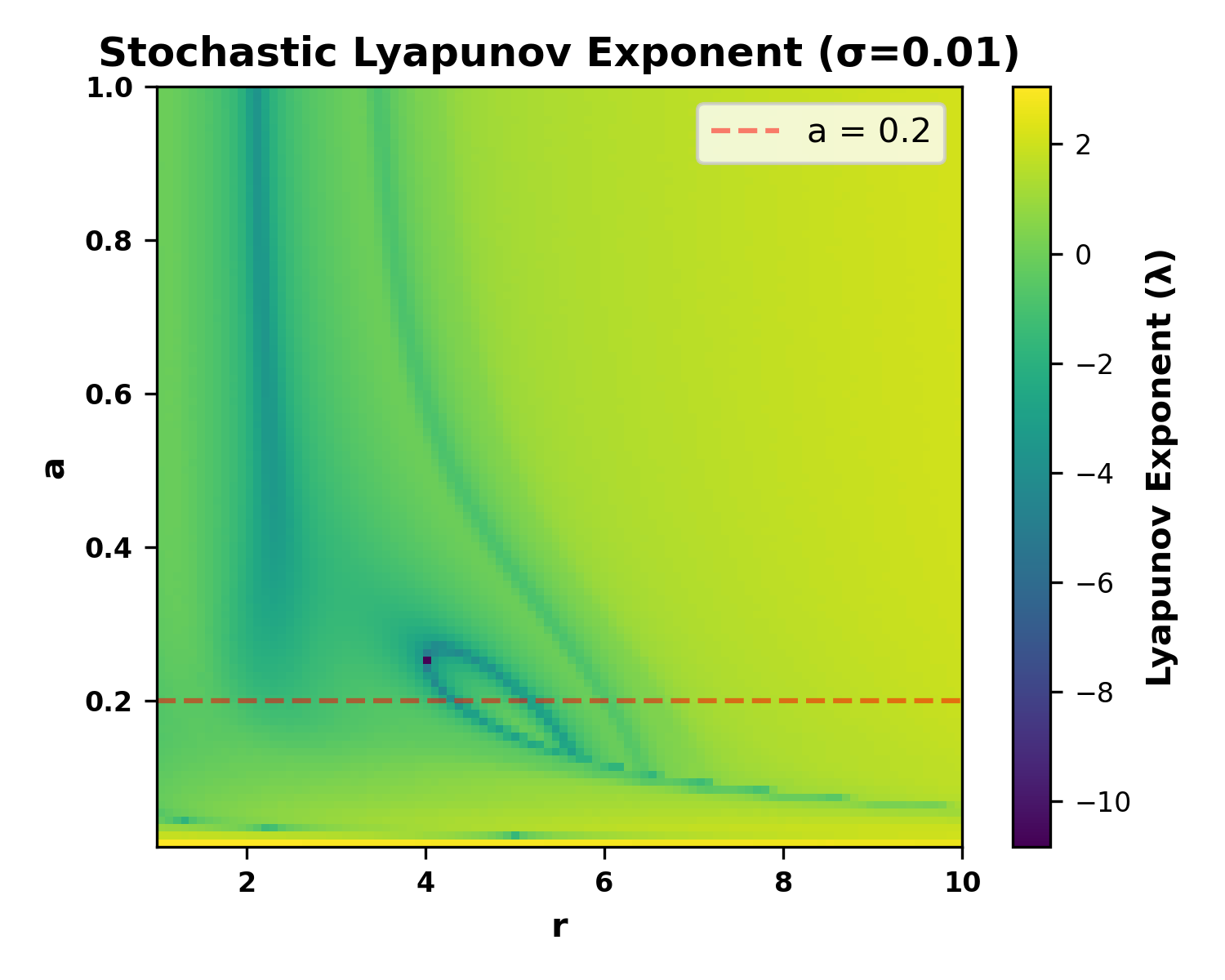}
   \caption{Stochastic Lyapunov spectrum in the $(r, a)$ parameter space. The system is perturbed by low-amplitude noise, which induces chaotic dynamics in regions that were previously associated with extinction in the deterministic case}
    \label{fig6}
\end{figure}

\begin{figure}[htbp]
    \centering
    \includegraphics[width=0.45\textwidth]{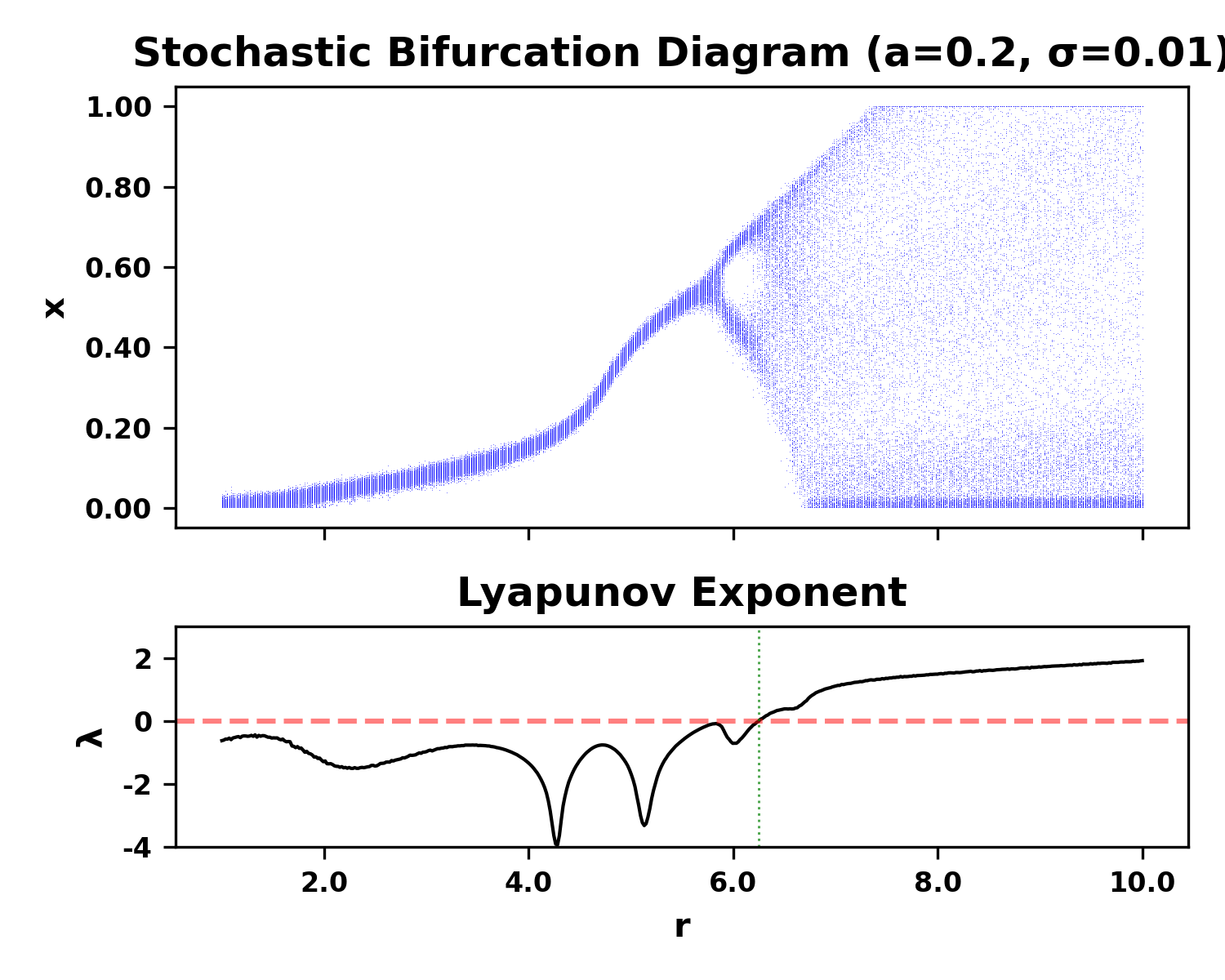}
   \caption{The bifurcation diagram and Lyapunov spectrum with noise are shown for a fixed value of $a = 0.2$ and a noise strength of 0.01.
}
    \label{fig7}
\end{figure}

We incorporate Gaussian noise (Equation\ref{eq3}) and generate the Lyapunov spectrum and bifurcation diagrams under stochastic forcing. Noise-induced transitions and stochastic bifurcations are identified and illustrated, as demonstrated in FIG~\ref{fig6},\ref{fig7}. 

For stochastic systems, the largest Lyapunov exponent (LLE) is computed along a noisy trajectory using:
\begin{equation}
\lambda = \lim_{N \to \infty} \frac{1}{N} \sum_{n=0}^{N-1} \ln \left| J(x_n) \right|,
\label{eqn17}
\end{equation}
where \( J(x_n) \) denotes the Jacobian evaluated along the trajectory at point \( x_n \), as defined by Equation~\ref{eqn11}.

The stochastic Lyapunov spectrum in the $(r, a)$ parameter space is shown in Fig.~(\ref{fig6}). The system dynamics are lightly perturbed by noise, which plays a significant role in inducing chaos. The dark blue and purple regions indicate stable behavior, where the system converges to fixed points or limit cycles. In contrast, the yellow and green regions exhibit noise-induced chaos.
A key observation is that the extinction regions present in the deterministic Lyapunov spectrum have now completely transformed into chaotic phases. In other words, the population is reinstated. Small fluctuations are sufficient to push the system out of inhibition traps, a phenomenon known as \textit{stochastic escape}.
To confirm the chaotic nature of the asymptotic states at higher parameter values, we have plotted the bifurcation diagram and Lyapunov exponent of the system for a fixed value of $a = 0.2$. These plots confirm the presence of stable chaotic oscillations in regions that were previously associated with extinction[see FIG~\ref{fig7}].

\begin{figure*}[htbp]
    \centering
   \includegraphics[width=0.9\textwidth, height=0.9\textheight]{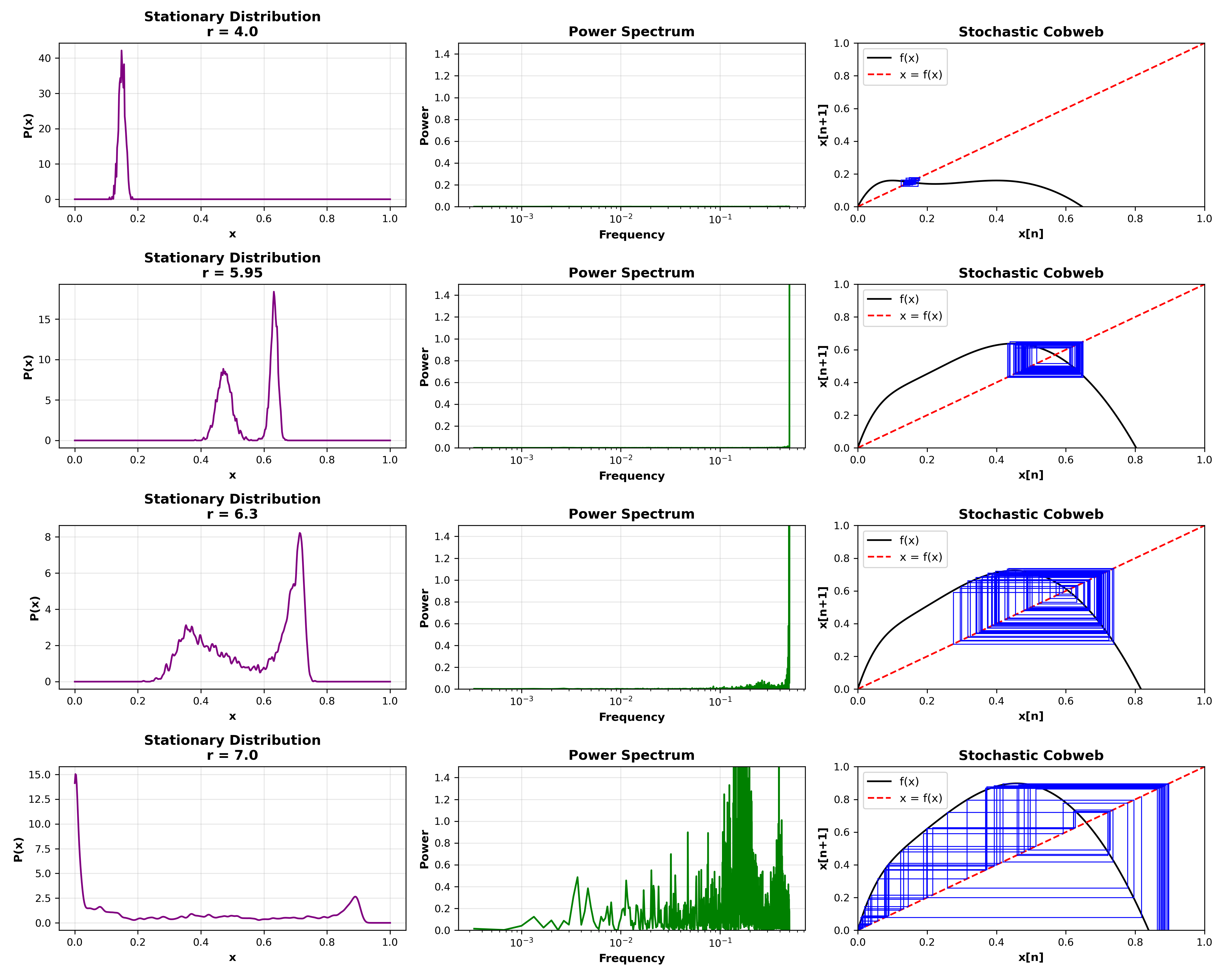}
    \caption{Stationary probability distribution(Left Panel), power spectral density(Middle Panel), and stochastic cobweb diagrams(Right Panel) for different values of the control parameter \( r \), highlighting noise-induced dynamics and transitions.}
    \label{fig8}
\end{figure*}

With the addition of noise, fixed points are no longer deterministic values but become \textit{probability distributions}. A stable fixed point in the stochastic system implies that the system fluctuates around a mean value, while the variance remains bounded. Unlike in the deterministic case, there is no sharp bifurcation point; instead, a \textit{probabilistic region} emerges where the system gradually changes its behavior, highlighting the inherent unpredictability of stochastic dynamics[see FIG~\ref{fig8}]. 
For the fixed value of \( r = 4 \), the sharply peaked unimodal distribution around \( x \approx 0.2 \) indicates that the system resides near a \textit{stochastically stable fixed point}. The corresponding power spectrum exhibits low activity and no distinct oscillatory behavior, supporting the interpretation of a stable, non-oscillatory regime. The cobweb diagram further confirms this by showing trajectories that converge steadily toward the fixed point.

At \( r = 5.95 \), a bimodal stationary distribution, a hallmark of bistability, is observed. This distribution, with its two distinct peaks, clearly indicates the presence of two preferred dynamical states. The power spectrum exhibits a pronounced peak, suggesting the existence of a period-two limit cycle. Notably, this periodic pattern persists even in the presence of noise, which causes the orbit to fluctuate around the deterministic cycle. This robustness under stochastic influence is a key characteristic of the system. The cobweb diagram further illustrates how iterates transition between states due to underlying metastability. The shape of the map suggests that the system is in a transitional regime, with dynamics influenced by both fixed points and periodic orbits. The system's proximity to a bifurcation point highlights a mixed dynamical regime, where small perturbations can shift the system between qualitatively different behaviours. 

When the parameter value is \( r = 6.3 \), the multimodal stationary distribution indicates the onset of weak aperiodic motion and the emergence of chaos—hallmarks of increasing dynamical complexity. The power spectrum reveals a broad frequency range with low-amplitude structure around a peak, reflecting the underlying irregular oscillations. The corresponding cobweb diagram exhibits a box-like density and scattered iterates, characteristic of a chaotic basin. Despite the presence of only weak noise, the system displays irregular and non-repeating trajectories, making the emergence of stochastic chaos clearly visible. For a higher value of \( r = 7.0 \), the stationary distribution becomes flatter and broader, with a dominant peak near \( x = 1 \). This indicates strong stochastic exploration of the state space. The power spectrum reveals a wide range of contributing frequencies, pointing to the presence of strong chaos. The cobweb diagram, with its visually striking and expanding web of iterates, vividly demonstrates how even weak noise can amplify dynamical complexity and enhance sensitivity to initial conditions. These features collectively reflect a transition into a fully chaotic regime.

The Lyapunov spectrum, bifurcation diagram, stationary distribution, power spectrum, and cobweb plots collectively reveal a clear transition from a stable fixed point, through bi-stability and metastability, to a chaotic regime. The presence of noise plays a crucial role in unmasking hidden attractors and resonances that remain inaccessible in the deterministic setting. These observations demonstrate that noise fundamentally reshapes the dynamical landscape, enabling unexpected persistence and facilitating the onset of complex, chaotic behaviour.

\subsection{Quasi-potential landscape}
The long-term behaviour of a stochastic system is governed not only by its deterministic dynamics but also by the influence of noise. To visualize and quantify the system's stability under external perturbations, we construct a \textit{quasi-potential} from time series data. This framework offers a global perspective on system stability and complements traditional tools such as Lyapunov exponents, time series analysis, and bifurcation diagrams. For a one-dimensional stochastic process with additive noise, as defined in Equation~(\ref{eq3}), the quasi-potential \( V(x) \) is derived from the stationary probability distribution \( P(x) \) of the system, according to the relation:

\begin{equation}
V(x) = -\log P(x)
\label{eqn18}
\end{equation}

where \( P(x) \) is obtained from the asymptotic distribution of the state variable \( x \). This formulation provides an energetic analogy of a particle rolling in a noisy landscape: \textit{local minima} correspond to preferred stable states, while \textit{barriers} and \textit{flat regions} represent metastable states, instabilities, or chaotic wandering. 
In this study, we employ the quasi-potential framework to investigate how weak noise reshapes the stability landscape of the underlying map. The computed quasi-potential \( V(x) \) reveals signatures of \textit{bi-stability}, \textit{suppression of extinction}, and \textit{restoration of chaotic dynamics} under weak noise. This approach captures the essential role of stochasticity and initial conditions in modifying the system's asymptotic behaviour, thereby highlighting the intricate and rich dynamics of noise-driven non-linear systems.

\begin{figure}[htbp]
    \centering
    \includegraphics[width=0.45\textwidth]{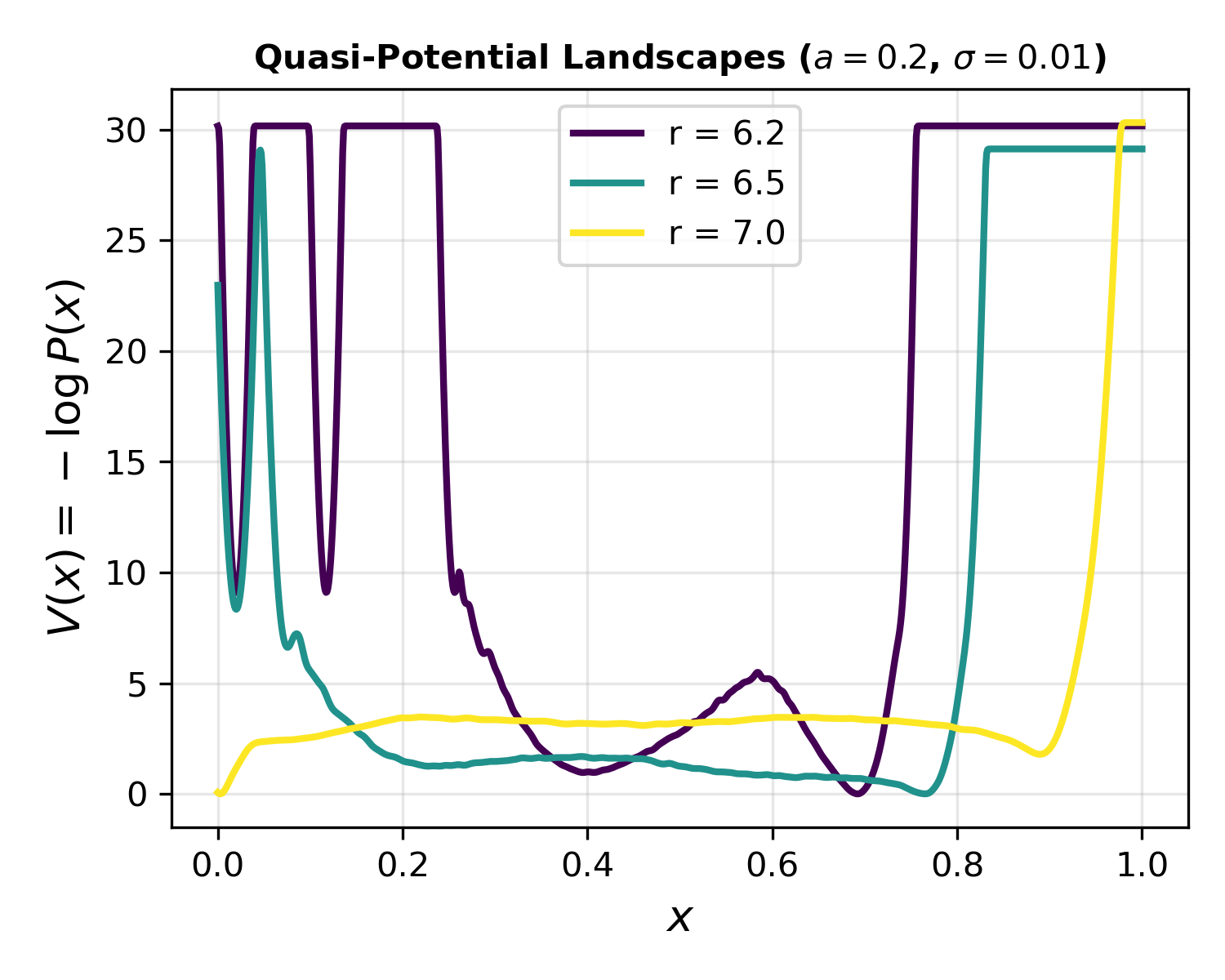}
    \caption{Quasi-periodic potential landscapes for \( r = 6.2 \) (violet) showing bi-stability, \( r = 6.5 \) (green) indicating onset of chaos, and \( r = 7 \) (yellow) depicting fully developed chaotic dynamics. The transition highlights the system's route from ordered bistable states to complex chaotic behaviour with increasing \( r \).}
    \label{fig9}
\end{figure}
From the quasi-potential calculation, we find that noise fundamentally alters the dynamical behaviour of the system, enabling transitions that are otherwise not permitted in the noise-free regime—such as escape from inhibitory traps leading to extinction. The plots of the quasi-potential \( V(x) \) reveal these transitions clearly [Fig.~\\ref{fig9}]. For example, at \( r = 6.2 \), the presence of sharp wells in the landscape indicates bi-stability(violet), allowing noise-induced switching between two metastable states. As \( r \) increases to 6.5(green) and 7.0(yellow), the potential landscape flattens significantly, suggesting the loss of dominant attractors and the onset of widespread chaotic exploration driven by stochasticity. 
Another crucial observation is the role of initial conditions. In deterministic systems, initial conditions determine the basin of attraction into which the system will settle. However, in the presence of noise, they influence how the system explores the quasi-potential landscape. For instance, under weak noise, initial conditions near a metastable state can push the system into trajectories that diverge and wander unpredictably—an effect clearly visible in the bifurcation diagrams and cobweb plots. This unpredictability under weak stochasticity is a surprising and insightful result of our study. It highlights the intricate interplay between noise and initial conditions in reshaping the stability structure and long-term behaviour of complex adaptive systems.
\section{\label{sec5}Result and Discussion}
The phenomenon highlighted by the hybrid model---\textit{deterministic collapse and stochastic recovery}---reveals a profound mechanism of \textbf{noise-induced stabilization}. In the absence of noise, the system exhibits an unstable saddle point or lies near the basin boundary at \( x_n = a \). Trajectories that approach this critical threshold tend to get trapped and eventually collapse to the extinction state \( x_n = 0 \). However, the presence of external perturbations \( \eta_n \) enables the system to escape this collapse basin and re-enter the stable manifold. This creates a clear \textit{stochastic rescue effect}, where noise counter intuitively prevents extinction by reactivating a dying system. These recovery paths are completely \textit{inaccessible in the deterministic regime}, yet under stochastic influence, the system exhibits resilience even near collapse. In this sense, noise acts as a \textbf{restorative force}, enabling survival where deterministic dynamics predict failure. This behaviour contrasts with the classical notion of noise-induced extinction and instead provides strong evidence for \textbf{noise-mediated persistence} in systems with \textit{threshold-triggered collapse}. The stochastic robustness observed here underscores the critical role of weak noise in maintaining system viability under inhibitory stress.\\
\\
Some real-world analogies, interpretations, and applications can be drawn from existing literature. One such immediate application involves the collapse of Harmful Algal Blooms (HABs). Suppose $x_n$ represents the algal biomass density, where a logistic term models nutrient-driven growth. A sigmoidal inhibition term captures auto-catalytic cell death due to toxin accumulation at high densities, triggering bloom collapse. This mechanism is observed in \textit{Microcystis} blooms, where a biomass threshold of approximately 20{,}000 cells/mL (corresponding to $a = 0.8$ in scaled units) marks the onset of collapse~\cite{ibelings2021harmful}. Now, if $\eta_n$ represents stochastic fluctuations in nutrient input or temperature, it can account for the resurgence of algal blooms following small rainfall events—scenarios where the deterministic model would otherwise predict irreversible collapse~\cite{tanaka2019ecolmod}. Another critical and relevant observation is that if \( x_n \) represents tumour cell counts—where the logistic growth term accounts for vascularisation-enabled proliferation, and the sigmoidal inhibition term captures immune system activation when \( x_n > a \)—this model aligns well with the phenomenon of dormancy escape in breast cancer (with \( a = 10^6 \) cells). The stochastic component (noise) represents variability in drug delivery or random immune cell infiltration, and it can explain noise-induced tumour reactivation and recurrence following therapy~\cite{robertsontessi2022natcomm}. This hybrid model has the potential to predict outcomes of checkpoint inhibitor therapy when combined with stochastic drug scheduling, thereby offering strategies to prevent relapse.\\
\\
Neuronal avalanches in cortical networks can be effectively described using this model by considering $x_n$ as the fraction of active neurons at time step $n$, with logistic growth representing excitatory propagation through the network. The sigmoidal inhibition term models the activation of inhibitory feedback mechanisms, which are triggered when activity surpasses a critical threshold, thereby preventing runaway excitation. Stochastic fluctuations, modeled as background synaptic noise, further shape the system's dynamics. Experimental observations in \textit{in vitro} cortical slice preparations reveal that such stochastic reactivation of neuronal avalanches occurs when the inhibition threshold emerges at approximately 25\% network activation ($a = 0.25$), highlighting the role of noise in sustaining near-critical dynamics in cortical circuits~\cite{priesemann2014jneurosci}.\\
\\
Social media information cascades~\cite{alvarez2023pnasnexus}, and fishery collapse and recovery~\cite{lindegren2016fish} are among the essential real-world applications of the proposed model. For any credible application, it is important to ensure that the threshold parameter, $a$, aligns with empirically known critical points. Additionally, the noise variance should reflect the actual environmental variability. For instance, in ecological systems, a realistic estimate is $\sigma^2 \approx 0.01\text{--}0.05$~\cite{lande2003stochastic}. Another key requirement is \textbf{bifurcation consistency}, meaning the collapse point must persist in the noise-free (deterministic) limit of the model.

\section{\label{sec6}Conclusion}
The proposed hybrid model offers a powerful framework for understanding critical transitions in complex systems. Its validity is supported through Lyapunov stability analysis, bifurcation diagrams, early warning signal (EWS) detection, and quasi-potential landscape reconstruction. This model stands out by capturing three universal features that are typically absent in classical approaches: (i) smooth threshold activation without artificial discontinuities, (ii) noise-induced recovery from deterministic collapse, and (iii) biologically meaningful upper bounds due to saturating inhibition. These properties reflect increasingly observed behaviors in both natural and social systems. By incorporating these mechanisms, the model moves beyond static tipping point theory, providing enhanced predictive capability and insights into dynamic resilience.\\
\\
Early warning signal (EWS) metrics—such as rising variance and autocorrelation—serve as useful indicators for anticipating tipping points. Quasi-potential analysis complements this by revealing how the system's stability landscape is reshaped under stochastic influences, illuminating transitions that can lead to population recovery even when deterministic dynamics predict collapse. By integrating both predictive indicators and a mechanistic understanding of collapse and recovery, the model proves especially well-suited for capturing the dynamics of systems with abrupt regulatory activation at critical thresholds. This makes it particularly applicable to domains such as social media cascades, ecological collapses, and neural feedback loops.\\
\\
In conclusion, by capturing critical thresholds and noise-induced transitions with both theoretical elegance and practical foresight, this model offers a robust framework for forecasting, managing, and potentially reversing collapse in real-world systems.
\bibliography{bibliography}
\end{document}